# DYNAMICS OF FASHION: THE CASE OF GIVEN NAMES

Damián H. Zanette

Centro Atómico Bariloche and Instituto Balseiro, 8400 Bariloche, Río Negro, Argentina

Email: zanette@cab.cnea.gov.ar

Abstract: We analyze the social mechanisms that shape the popularity rise and fall of the names given to newborn babies. During the initial stage, popularity increases by imitation. As the people with the same name grow in number, however, its usage is inhibited and eventually decays. This process mirrors the dynamics of fashion fads. An activator-inhibitor dynamical model for the interplay of the population bearing a name and the expecting couples wishing to give it to their children provides a satisfactory explanation of historical data from the Canadian province of Quebec during the twentieth century.

Within human social groups, individuals adopt common cultural traits as a way to strengthen the identity of their own group and enhance differences with other communities [1]. Inside a given group, however, cultural variations are needed to maintain the individuality of each member. This duality engenders opposite forces in the evolution of culture. In particular, it drives the recursive dynamics of adoption and abandonment of the habits, styles, and manners that we associate with fashion.

The name given to a newborn baby, usually chosen by the parents, is a cultural trait. As such, the usage of given names is subjected to the pressures that shape changes in culture. The frequency with which different names are used over the years follows strikingly uniform patterns. Of the two hundred most common names given to children during the twentieth century in the Canadian province of Quebec, for instance, more than two thirds showed a period of rapid growth in their usage frequency followed by a comparatively slow decay (Fig. 1A). At its maximum, the frequency typically attained a few percent of births, with some extreme cases getting above 10 %. Altogether, growth and decay lasted for a few decades, after which the frequency returned to its initial level, fluctuating below 0.1 % [2]. Similar patterns have been reported in the United States and in France [3]. Over the relatively long time scales of demographic change, this behavior mirrors the popularity rise and decline of fads and crazes.

The Quebec data suggests that, for most names, there is at least one expecting couple out of ten thousand who will give one of them to their baby, irrespectively of the others' choice. This explains the small but sustained usage frequency of most names before their popularity begins to rise. An upwards fluctuation may however make the name conspicuous enough as to trigger imitation, with the ensuing increase of its frequency. The imitation process is initially autocatalytic [4]: the larger the number of parents giving the name, the higher its growth rate. As the number of young people with the same name grows, however, the perception that such a widespread name may endanger their child's (and their own) identity within their social group dissuades more and more expecting parents from giving it. Eventually, the usage frequency reaches a maximum and, since the population bearing the name is still growing, it later decreases towards its initial level.

To give a quantitative description of these mechanisms, we consider the dynamic interplay between two parts of the population: the people who bear the name, and the number of couples which give the name to their newborn children. The growth rate of the first group is given by the difference between their birth rate, which is proportional to the size of the second group, and the death rate. The second group, in turn, must overcome a certain threshold size for imitation to act, above which it grows at a rate proportional to its own size. This rate, however, decreases as the first group grows. For larger sizes, moreover, also the second group inhibits its own growth. Details of the mathematical implementation of this dynamical model are given in the supporting material. As illustrated in Fig. 1B, the model provides very good fittings of the observed evolution of the usage frequency of given names. We obtained fittings of similar quality for much of the Quebec data. Besides reproducing the asymmetric duration of frequency growth and decrease, the model predicts that the overall evolution is faster when the popularity peak is higher, as actually observed to happen. It would be interesting to advance an explanation, not provided by the model, for the fact that the popularity change seems to have become sharper towards the end of the twentieth century.

Our model for the rise and decline of given names belongs to the class of activator-inhibitor dynamical systems, which have been applied to the description of biological morphogenesis, disease spreading, and nerve impulses [5]. In this perspective, human social groups are seen as excitable entities, prone to recurrently stimulate and inhibit themselves through the waves of fashion.

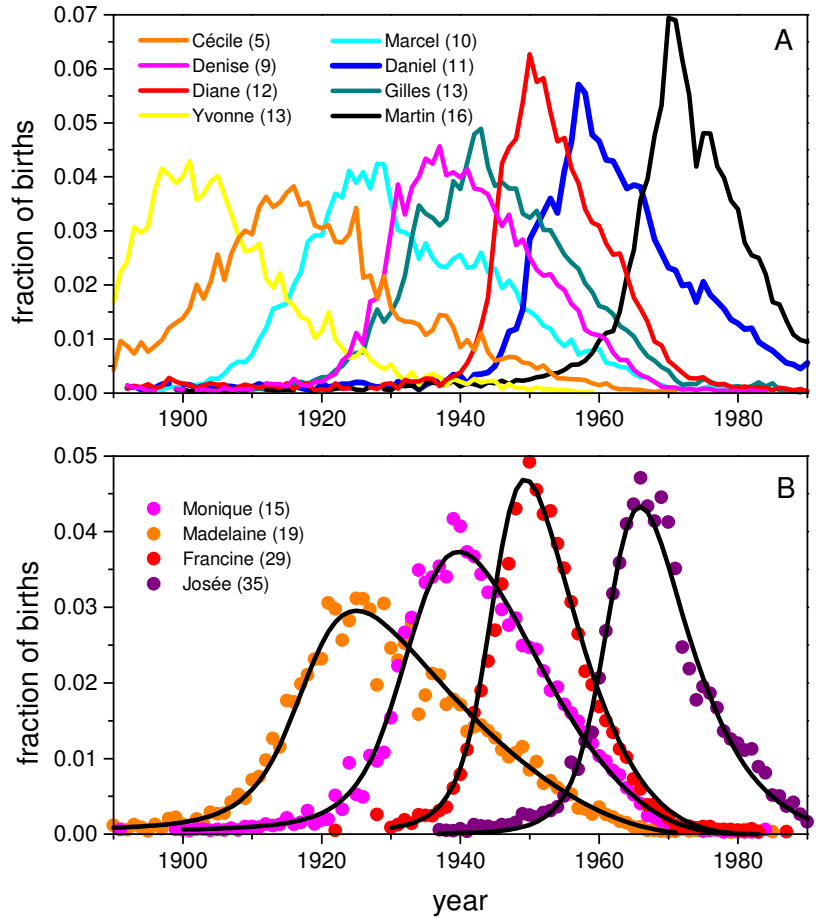

Fig. 1. (A) Fraction of newborn babies given a specific name in the period 1890-1990 in the Canadian province of Quebec, for four female and four male names. Numbers in brackets give the overall usage rank of each name, discerning between the two genders, in the period 1889-2000. (B) Fittings of the fraction of newborn babies (black curves) for four female names from Quebec (not included in panel A). The mathematical model describes initial imitation between expecting parents, as well as subsequent inhibition due to the perception that a given name is becoming too frequent.

SUPPORTING MATERIAL

The social mechanisms that determine the popularity rise and decline of a given name are modeled by means of an activator-inhibitor dynamical system [1]. Let $u_t$ and $v_t$ respectively be the fraction of couples that give that name to their newborn child (activator) and the fraction of the population that bears the name (inhibitor), both in year $t$. The annual increment of $v_t$ is

$$\Delta v_t = v_{t+1} - v_t = u_t - \mu_t v_t, \qquad (1)$$

with $\mu_t$ the annual mortality in year $t$. This gives the difference between birth and death events. For $u_t$ we propose

$$\Delta u_t = u_{t+1} - u_t = -\alpha u_t^p + \beta(u_t, v_t) u_t. \qquad (2)$$

For $p<1$, the first term in right-hand side of Eq. (2) determines a threshold in the growth rate of $u_t$, at $u \approx [\alpha/\beta(0,0)]^{1/(1-p)}$. Above this threshold, the growth of $u_t$ is autocatalytic, at rate $\beta(u_t, v_t) = \beta(0,0) (1 - u_t / u_S) (1 - v_t / v_S)$. For small $u_t$ and $v_t$, this growth rate is approximately constant: $\beta(u_t, v_t) \approx \beta(0,0)$. As either $u_t$ and $v_t$ grow, however, $\beta(u_t, v_t)$ declines, and vanishes at the saturation values $u_t = u_S$ and $v_t = v_S$. Above these values the growth rate of the fraction of couples is negative.

We have dealt with Eqs. (1) and (2) in their continuous-time version, where they become differential equations. For specified values of the mortality $\mu_t$, Eq. (1) can be formally solved to give $v_t$ in terms of $u_t$. This solution is then replaced into Eq. (2) to get an integro-differential equation for $u_t$. The empirical data for $u_t$ corresponding to specific given names, which coincide with the fraction of births illustrated in Fig. 1 of the main text, were first fitted using a log-polynomial continuous approximation. Then, we have calculated the annual increments in $u_t$ from the fitting. The parameters $\alpha$, $p$, $\beta(0,0)$, $u_S$, and $v_S$ have been estimated by nonlinear regression between $u_t$ and $\Delta u_t$. Finally, using these parameter values, Eq. (2) has been integrated. The curves in Fig. 1B of the main text are the result of the integration.

The empirical data of usage frequency of given names in Quebec were compiled and provided by L. Duchesne, from *Institut de la statistique du Québec*, who had previously published an elaboration and analysis of the same data [2]. Quebec mortality rates for the period 1890-1990 were obtained from the webpage of the *Institut de la statistique du Québec*, www.stat.gouv.qc.ca.